\documentclass[english]{article}
\usepackage[T1]{fontenc}
\usepackage[latin1]{inputenc}
\usepackage{geometry}
\usepackage{array}
\usepackage{prettyref}
\usepackage{float}
\usepackage{amssymb}

\makeatletter

\providecommand{\LyX}{L\kern-.1667em\lower.25em\hbox{Y}\kern-.125emX\@}
\newcommand{\noun}[1]{\textsc{#1}}

 \usepackage{verbatim}

\newcommand{\td}[2]{\mathcal{D}_{#1}\left( #2 \right) }
\newcommand{\tdn}[1]{\mathcal{D}_{#1} }

\usepackage{babel}
\makeatother
\begin{document}

\title{Generalized Verlinde formulae for some Riemann surface automorphisms}

\author{Tamas Varga\\
{\small Department of Theoretical Physics, Roland Eötvös University,
1117 Budapest, Hungary} \\
{\small E-mail: vargat@elte.hu}}

\maketitle
\begin{abstract}
The generalized Verlinde formulae expressing traces of mapping classes
corresponding to automorphisms of certain Riemann surfaces, and the
congruence relations on allowed modular representations following
from them are presented. The surfaces considered are families of algebraic
curves given by suitably chosen equations, the modular curve $\mathcal{X}(11)$,
and a factor curve of $\mathcal{X}(8)$. The examples of modular curves
illustrate how the study of arithmetic properties of suitable modular
representations can be used to gain information on automorphic properties
of Riemann surfaces.
\end{abstract}

Rational Conformal Field Theories (RCFT) \cite{cft} appear in several
areas of physics and mathematics. Their classification would imply
important results in e.g. string theory, 2D critical phenomena and
topology of 3D manifolds. A key part of the classification problem
is the classification of allowed modular representations corresponding
to a RCFT. This representation is generated by the matrices $S$ and
$T$ that are determined by modular properties of genus 1 characters:

\begin{eqnarray}
\chi _{p}(\frac{-1}{\tau }) & = & \sum _{q}S_{pq}\chi _{q}(\tau )\label{eq:modrep}\\
\chi _{p}(\tau +1) & = & \sum _{q}T_{pq}\chi _{q}(\tau )\nonumber 
\end{eqnarray}
Besides the modular relations $S^{4}=\mathbb{I}$, $(ST)^{3}=S^{2}$,
which ensure that they generate the group $SL(2,\mathbb{Z})$, they
have to satisfy additional constraints coming from the extra structure
of a RCFT lying behind them (see e.g. \cite{gannon}). A well known example of these is the
formula of Verlinde\begin{equation}
\textrm{dim}\mathcal{V}_{0}(p,q,r)=\sum _{s}\frac{S_{ps}S_{qs}S_{rs}}{S_{0s}}\label{eq:Verlinde}\end{equation}
which expresses the dimension of the space of genus $0$ holomorphic
blocks with insertions of primaries $p,q,r$ in terms of modular matrix
elements. Clearly, for all allowed modular representations the above
combination of matrix elements has to give a positive integer. But
Verlinde's theorem alone doesn't give a sufficient condition for a
modular representation to come from a RCFT. For example, it was found
in \cite{bantvecs} that it is possible to obtain independent constraints
by generalizations of Verlinde's formula. The complete characterization
of these formulae and their geometric origin was presented in \cite{genverlinde}
- in the following we shall review its main results. 

It follows from Verlinde's formula \prettyref{eq:Verlinde} that the
dimension of the space of genus $g$ holomorphic blocks (denoted as
$\mathcal{V}_{g}$) might be expressed as \cite{mooreseiberg}\begin{equation}
\textrm{dim}\mathcal{V}_{g}=\sum _{q}S_{0q}^{2-2g}\label{eq:verlindevg}\end{equation}
In a similar way as the modular group is represented on $\mathcal{V}_{1}$,
a RCFT defines a natural representation of the mapping class group
$M_{g}$ on the space $\mathcal{V}_{g}$. (The elements of $M_{g}$
are equivalence classes of conformal mappings of genus $g$ surfaces
generated by Dehn twists. These act on the Teichmüller space $\mathbb{T}_{g}$
of genus $g$ closed Riemann surfaces and are also projectively represented
on the space of genus $g$ holomorphic blocks - a special case of
this for $g=1$ is the action of the modular group on the upper half-plane,
giving the modular representation \prettyref{eq:modrep}.) The formula
\prettyref{eq:verlindevg} expresses the trace of the identity operator
of these representations. The main result of \cite{genverlinde} was
to show that the traces of all finite order elements of $M_{g}$ might
be expressed via generalized Verlinde formulae in terms of modular
matrix elements. 

In order to write generalized Verlinde formulae in a compact form, we first have to define $\Lambda$-matrices.  For $r \in \mathbb{Q}$ take any $M \in SL(2,\mathbb{Z})$ such that $r=M(\infty)$, let $-r^{*}=M^{-1}(\infty)$, and then $\Lambda(r)=T^{r}MT^{r^*}$ ($M$ and $T$ denote here the representation matrices on $\mathcal{V}_1$). It can be shown that this definition is independent of the choice of $M$, and that $\Lambda$-matrices have the following properties: $\Lambda(r+1)=\Lambda(r)$ and $\Lambda(0)=S$. 

The appropriate generalizations of the right hand side of \prettyref{eq:verlindevg} are the twisted dimensions defined as  

\[
\td {g}{r_{1}\ldots r_{n}}=\sum _{q}S_{0q}^{2-2g}\prod _{i=1}^{n}\frac{\Lambda _{q0}(r_{i})}{S_{0q}}\quad (r_{1},\ldots r_{n}\in \mathbb{Q})\]
The rationals $(r_{1}\ldots r_{n})$
are called the characteristics of the twisted dimension. We shall
later use the shorter notation $\mathcal{D}_{g}\left([r_{1}]^{l_{1}},\ldots ,[r_{n}]^{l_{n}}\right)$
for the twisted dimension whose characteristics contains the rational
$r_{i}$ $l_{i}$ times. Twisted dimensions and $\Lambda$-matrices play an important role in the theory of permutation orbifolds, and more about their properties can be found in \cite{permorb}.  The name ''twisted dimension'' expresses the fact that, with the above notation, \prettyref{eq:verlindevg} might be written as

\[
\textrm{dim}\mathcal{V}_{g}=\tdn {g}\]

The trace formula given in \cite{genverlinde} for $\gamma \in M_{g}$
is\begin{equation}
Tr(\gamma )=\mathcal{D}_{g*}\left(r_{1},\ldots ,r_{n}\right)\label{eq:genverlinde0}\end{equation}
where $g^{*}$ and $r_{1},\ldots ,r_{n}$ are determined by $\gamma $
in the following way. It is known from a theorem of Nielsen \cite{nielsen}
that if $\gamma $ is of finite order, then it has a fixed point in
its action on Teichmüller space. Furthermore, $\gamma $ might be
lifted to an automorphism (i. e. a biholomorphic self-mapping) of
the Riemann surface corresponding to the fixed point. The parameters
$g^{*}$ and $r_{1},\ldots ,r_{n}$ in \prettyref{eq:genverlinde0}
are certain characteristics of this automorphism (we shall return
to the details later). This approach was taken in \cite{genverlinde}
to obtain trace formulae for finite order elements of $M_{1}$. The
finite order elements of this group are (conjugate to) members of
the subgroups $\left\langle S\right\rangle $ and $\left\langle ST\right\rangle $
that fix the points $\tau =i,e^{\frac{2\pi i}{3}}$, respectively.
Unfortunately, much less is known about finite order elements of higher
genus mapping class groups and their fixed points in Teichmüller space.
These correspond to Riemann surfaces with non-trivial automorphism
groups, and the theory of these (especially the ones with {}``large''
automorphism groups) is a rich area of mathematics with many open
questions \cite{macbeath}. One of the basic theorems here is a theorem
of Hurwitz, which states that the order of the automorphism group
of a genus $g\ge 2$ surface is bounded by $|\mathcal{A}|\le 84(g-1)$.
The famous lowest genus example where this bound is saturated is the
Klein quartic - a genus 3 surface with automorphism group $PSL(2,7)$
of order 168. 

Even if a systematic treatment of surfaces, like in the case of $M_{1}$,
is not possible, we may pick some Riemann surfaces with well known
automorphic properties, and try to write down the trace formulae corresponding
to their automorphisms. In \cite{genverlinde} this was illustrated
on the example of the Klein quartic. The aim of the present note is
to consider other Riemann surfaces where a similar analysis can be
carried out. First, we consider families of algebraic curves given
by algebraic equations that are suitable for our purposes (i.e. some
of their automorphisms can be given in a particularly simple form).
This will lead to infinite series of trace formulae. After that, we
shall turn to modular curves, mainly because they are interesting
from a mathematical point of view - e.g. they possess {}``large''
automorphism groups. The arising trace formulae are presented for
two surfaces - that is the full character of their automorphism group
(of order 84 and 660) is given in terms of twisted dimensions.

Let's now recall how an automorphism of a Riemann surface determines
the characteristics of the twisted dimension in the corresponding
trace formula. Consider a closed genus $g$ Riemann surface $\mathcal{S}$
and an element of its automorphism group $\gamma \in Aut(\mathcal{S})$
of order $N$. One may construct a new surface $\mathcal{S}/\gamma $
by identifying the points on $\gamma $ orbits. The complex structure
of $\mathcal{S}/\gamma $ is defined so that the natural projection
map $\pi :\mathcal{S}\to \mathcal{S}/\gamma $ becomes a holomorphic
$N$ sheeted covering map. The parameter $g^{*}$ in \prettyref{eq:genverlinde0}
denotes the genus of $\mathcal{S}/\gamma $. The points on $\gamma $
orbits with non-trivial stabilizer subgroups correspond to ramification
points of $\pi $, and the ramification index equals the order of
the stabilizer subgroup. Let the number of such $\gamma $ orbits
be $r$, and the order of stabilizer subgroups be $n_{1},\ldots ,n_{r}$
(clearly $n_{i}|N$). There are $m_{i}=N/n_{i}$ ramification points
on each orbit, so the total branching order of $\pi $ is $B=\sum _{i=1}^{r}\frac{(n_{i}-1)N}{n_{i}}$.
Then the genus $g^{*}$ might be expressed by the Riemann-Hurwitz
formula as

\begin{equation}
g^{*}=\frac{g-1}{N}-\sum _{i=1}^{r}\frac{(n_{i}-1)}{2n_{i}}+1\label{eq:hurwitz}\end{equation}
$(g^{*};n_{1},\ldots ,n_{r})$ shall be referred to as the signature
of $\gamma $ (or the corresponding branched covering). In terms of
a suitably chosen local coordinate $z$ the action of $\gamma ^{N/n_{i}}$
(which is an element of the stabilizer subgroup) can be written near
a branch point on the $i$-th orbit as

\begin{equation}
\gamma ^{N/n_{i}}:z\to e^{2\pi i\frac{k_{i}}{n_{i}}}z\label{eq:monodromydef}\end{equation}
where $0< k_{i}< n_{i}$, and is coprime to $n_{i}$. The rationals
in the trace formula \prettyref{eq:genverlinde0} are then defined
as $r_{i}=\frac{k_{i}}{n_{i}}$, and they are called the monodromy
of the covering. 

It will be useful to notice a few consequences of such trace formulae
on the arithmetic properties of the involved twisted dimensions. First,
let's note that $\gamma ^{l}$ for $gcd(l,N)=1$ is also an order
$N$ automorphism of the surface, with the same orbits as $\gamma $.
Its signature is therefore the same as that of $\gamma $, and further,
from the definition \prettyref{eq:monodromydef} it follows that its
monodromies are $(lk_{i})/n_{i}$, where we use the natural convention
that multiplication (and inversion) of $k_{i}$-s should be understood
mod $n_{i}$. Thus the trace of $\gamma ^{l}$ is expressed by the
formula

\begin{equation}
Tr(\gamma ^{l})=\mathcal{D}_{g*}\left(\frac{lk_{1}}{n_{1}},\ldots ,\frac{lk_{r}}{n_{r}}\right)\quad \textrm{if }gcd(l,N)=1\label{eq:gammaexp}\end{equation}
If $gcd(l,N)>1$, then the signature and monodromies of $\gamma ^{l}$
are still determined by those of $\gamma $, but the rule is more
complicated - a few examples can be read off from Tables \ref{cap:psl11},\ref{cap:psl8/z2}.
In the most important case when $N$ is prime, this means that $Tr_{V_{g}}(\mathbb{I})=\mathcal{D}_{g}$
and $Tr(\gamma ^{l})=\mathcal{D}_{g*}\left(\frac{lk_{1}}{N},\ldots ,\frac{lk_{r}}{N}\right)$
for $1\le l<N$ form a character of the group $\left\langle \gamma \right\rangle \cong \mathbb{Z}_{N}=\mathbb{Z}/N\mathbb{Z}$.
If $N$ is prime, this is equivalent to the following conditions on
the twisted dimensions (which they should satisfy in any RCFT):

\begin{enumerate}
\item The fact that the scalar product of this character with the trivial
character of $\mathbb{Z}_{N}$ is a multiple of $N$, implies the
following congruence relation on twisted dimensions\begin{equation}
\mathcal{D}_{g}+\sum _{l=1}^{N-1}\mathcal{D}_{g*}\left(\frac{lk_{1}}{N},\ldots ,\frac{lk_{r}}{N}\right)\equiv 0\quad \textrm{mod }N\label{eq:congrel}\end{equation}

\item The formula \prettyref{eq:genverlinde0} also restricts the Galois
transformation properties of twisted dimensions. It is known that
$Tr(\gamma )$ is an element of $\mathbb{Q}[\zeta _{N}]$, where $\zeta _{N}=exp(\frac{2\pi i}{N})$
($N=ord(\gamma )$ is not necessarily a prime here). The Galois group
Gal$(\mathbb{Q}[\zeta _{N}]/\mathbb{Q})$ acts on this field by Frobenius
transformations $\sigma _{l}$ ($gcd(l,N)=1$), which leave $\mathbb{Q}$
fixed and take $\zeta _{N}$ to $\zeta _{N}^{l}$. It is clear that
$Tr(\gamma ^{l})=\sigma _{l}(Tr(\gamma ))$, therefore the twisted
dimensions appearing in trace formulae possess the following Galois
transformation property\begin{equation}
\sigma _{l}\left(\mathcal{D}_{g*}\left(\frac{k_{1}}{n_{1}},\ldots ,\frac{k_{r}}{n_{r}}\right)\right)=\mathcal{D}_{g*}\left(\frac{lk_{1}}{n_{1}},\ldots ,\frac{lk_{r}}{n_{r}}\right)\quad \textrm{if }gcd(l,N)=1\label{eq:galoistrans}\end{equation}
The Galois action on $\Lambda $ matrices was already studied in \cite{galoistrans}
using permutation orbifold techniques. The same Galois transformation
property of twisted dimensions follows from the results of \cite{galoistrans}
if $N$ is prime, and $\sum _{i=1}^{r}\frac{k_{i}^{-1}}{N}\in \mathbb{Z}$,
where $k_{i}^{-1}$ is the mod $N$ inverse of $k_{i}$. The later
condition is always satisfied by monodromies as a consequence of the
monodromy theorem (we shall show this explicitly in the case of some
algebraic curves). Therefore, if $N$ is prime, the only new constraint
on twisted dimensions is the congruence relation \prettyref{eq:congrel}. 
\end{enumerate}
It's possible to strengthen these restrictions if $\gamma ^{l}$ is
conjugate to $\gamma $ in $M_{g}$ for all $l\in M$ where $M$ is
a multiplicative subgroup of $\mathbb{Z}_{N}^{*}$. Then $\mathcal{D}_{g*}\left(\frac{k_{1}}{n_{1}},\ldots ,\frac{k_{r}}{n_{r}}\right)$
has to be invariant under the action of $\sigma _{l}$ if $l\in M$,
therefore is restricted in all RCFT-s to fall in the field $\mathbb{Q}[\sum _{l\in M}\zeta _{N}^{l}]$.
This also allows us to simplify the congrunce relations. In particular,
if $N$ is prime and $M=\mathbb{Z}_{N}$, then $\mathcal{D}_{g*}\left(\frac{k_{1}}{n_{1}},\ldots ,\frac{k_{r}}{n_{r}}\right)$
is invariant under the full Galois group, therefore is an integer
(because it is a trace). Then the congruence relation \prettyref{eq:congrel}
can be written, after adding $\mathcal{D}_{g*}\left(\frac{k_{1}}{n_{1}},\ldots ,\frac{k_{r}}{n_{r}}\right)$
to both sides, as $\mathcal{D}_{g*}\left(\frac{k_{1}}{n_{1}},\ldots ,\frac{k_{r}}{n_{r}}\right)\equiv \mathcal{D}_{g}$.

After having seen how generalized Verlinde formulae restrict the arithmetic
properties of twisted dimensions, let us now evaluate them for some
explicit Riemann surface automorphisms. In the first example we shall
construct automorphisms whose signatures and monodromies can be easily
read off from their definition. 

Consider first the surface $\mathcal{S}$ given by the equation

\begin{equation}
w^{N}=(z-e_{1})^{k_{1}}\ldots (z-e_{r})^{k_{r}}\label{eq:algcurv1}\end{equation}
where $e_{i}$ $(i=1\ldots r)$ are distinct points in $\mathbb{C}$,
$1\le k_{i}\le N$ are coprime to $N$, and $\sum _{i=1}^{r}(k_{i}/N)\in \mathbb{N}$.
In this case the mapping $z:\mathcal{S}\to \mathbb{P}^{1}$ is an
$N$ sheeted branched covering map (the points over $z$ are $(\zeta _{N}^{i}w,z)$
$i=0\ldots N-1$ for any solution $w$ of \prettyref{eq:algcurv1}).
The ramification points of ramification index $N$ are $(0,e_{i})$,
$i=1\ldots r$.%
\footnote{To see this, first choose a local coordinate $z_{i}(z)$ in the neighborhood
of $e_{i}$ such that \begin{equation}
z_{i}(e_{i})=0\quad \textrm{and}\quad (z-e_{1})^{k_{1}}\ldots (z-e_{r})^{k_{r}}=z_{i}^{k_{i}}\label{eq:coordzi}\end{equation}
 where the second equation has a locally biholomorphic solution, because
both sides are of order $k_{i}$ at $e_{i}$. In the new coordinate
the Puisseux series over $z_{i}=0$ is $w=z_{i}^{k_{i}/N}$, which
is of ramification index $N$ since $k_{i}$ is coprime to $N$. We
also need to prove that $\mathcal{S}$ is unbranched over $z=\infty $.
For this choose a local coordinate $z_{\infty }(\infty )=0$ and $(z-e_{1})^{k_{1}}\ldots (z-e_{r})^{k_{r}}=z_{\infty }^{-\sum _{i=1}^{r}k_{i}}$,
which is again possible because the order of the left hand side at
$\infty $ is $-\sum _{i=1}^{r}k_{i}$. In this coordinate the function
elements over $\infty $ are $w=\zeta _{N}z_{\infty }^{-\sum _{i=1}^{r}(k_{i}/N)}$,
and these are unbranched because $\sum _{i=1}^{r}(k_{i}/N)\in \mathbb{N}$.
This explains our choice $\sum _{i=1}^{r}(k_{i}/N)\in \mathbb{N}$
- if we wouldn't require it, the cover would be branched over $\infty $,
and the same condition would hold after including the monodromy at
$\infty $. In a topological approach this might be seen to be the
consequence of the monodromy theorem. %
} The total branching order of $z$ is then $B=(N-1)r$ ($B$ is even
because either $N$ or $k_{i}$-s are odd, but in the later case $r$
has to be even if $\sum _{i=1}^{r}(k_{i}/N)\in \mathbb{N}$), therefore
the genus of $\mathcal{S}$ is $g=B/2-N+1=(N-1)(r/2-1)$ using \prettyref{eq:hurwitz}.
The covering transformation $t:(w,z)\mapsto (\zeta _{N}w,z)$ is an
automorphism of order $N$, whose fixed points are the ramification
points of the cover. The orbits of $t$ are the points over $z$,
therefore the natural projection map $\pi _{t}:\mathcal{S}\to \mathcal{S}/t$
is the same mapping as $z$. This means that, since $\mathcal{S}/t\cong \mathbb{P}^{1}$,
the signature of $t$ is $(0;N,\ldots ,N)$ where $N$ appears $r$
times. In order to determine the monodromy of $t$ at a given point
$(0,e_{i})$, we shall choose a suitable local coordinate on $\mathcal{S}$
near this point. Since the mapping $z_{i}(z):\mathcal{S}\to \mathbb{C}$
(see \prettyref{eq:coordzi}) defined in a neighborhood of $(0,e_{i})$
is a locally holomorphic mapping of ramification index $N$ in $(0,e_{i})$
(because $z$ is of ramification index $N$, and $z_{i}$ is locally
biholomorphic), and takes $(0,e_{i})$ to $0$, there exists a local
coordinate $\lambda $ on $\mathcal{S}$ near $(0,e_{i})$, such that
$z_{i}(z)=\lambda ^{N}$. Since $t$ leaves $z$, thus also $z_{i}(z)$,
invariant, it has to act on $\lambda $ as $t:\lambda \mapsto \zeta _{N}^{l}\lambda $
for some $0\le l\le N-1$. From \prettyref{eq:coordzi} and \prettyref{eq:algcurv1}
it follows that $w^{N}=z_{i}^{k_{i}}$, therefore we can choose $\lambda $,
such that $w=\lambda ^{k_{i}}$. Since $\zeta _{N}w=t(w)=t(\lambda )^{k_{i}}=\zeta _{N}^{lk_{i}}\lambda ^{k_{i}}=\zeta _{N}^{lk_{i}}w$
it follows that $l\equiv k_{i}^{-1}$ mod $N$. Therefore, we have found
that the monodromy of $t$ at $(0,e_{i})$ is $k_{i}^{-1}/N$, thus
the generalized Verlinde formula for $t$ is the following

\[
Tr(t)=\td {0}{\frac{k_{1}^{-1}}{N},\ldots ,\frac{k_{r}^{-1}}{N}}\]
This yields the following congruence relation (see \prettyref{eq:congrel},
and recall that the genus of $\mathcal{S}$ is $g=(N-1)(r/2-1)$)
\begin{equation}
\mathcal{D}_{(N-1)(\frac{r}{2}-1)}+\sum _{l=1}^{N-1}\mathcal{D}_{0}\left(\frac{lk_{1}^{-1}}{N},\ldots ,\frac{lk_{r}^{-1}}{N}\right)\equiv 0\quad \textrm{mod }N\label{eq:congrel-alg}\end{equation}
for any choice of $1\le k_{i}\le N-1$ coprime to $N$ such that $\sum _{i=1}^{r}(k_{i}/N)\in \mathbb{N}$.
Further, we obtained that under Galois transformations $\mathcal{D}_{0}\left(\frac{k_{1}^{-1}}{N},\ldots ,\frac{k_{r}^{-1}}{N}\right)$
behaves as in \prettyref{eq:galoistrans} which is in accordance with
the results of \cite{galoistrans} since $\sum _{i=1}^{r}(k_{i}/N)\in \mathbb{N}$.
A similar, but more lengthy analysis could be done in the case if
$k_{i}$ is not necessarily relatively prime to $N$. The main difference
is that there is in general more than one point over $z=e_{i}$, therefore
one has to be more careful in determining the action of $t$ on (and
near) these orbits. 

To construct an algebraic curve with more automorphisms, consider
a surface $\mathcal{S}'$ similar to the example in \cite{farkas-kra}
p. 247

\begin{equation}
w^{N}=(z^{q}-e_{1}^{q})^{k_{1}}\ldots (z^{q}-e_{r}^{q})^{k_{r}}\label{eq:algcurv2}\end{equation}
where $0\ne e_{i}\in \mathbb{C}$, $1\le k_{i}\le N-1$, $\sum _{i=1}^{r}(k_{i}/N)\in \mathbb{N}$.
For simplicity, let both $N\ne q$ be primes. Note that this is a
special case of \prettyref{eq:algcurv1} since $(z^{q}-e_{i}^{q})=\prod _{j=0}^{q-1}(z-\zeta _{q}^{j}e_{i})$,
thus in this case the ramification points of $proj$ (all of ramification
index $N$) are the $qr$ points $\zeta _{q}^{j}e_{i}$ ($j=0\ldots q-1$,
$i=1\ldots r$). Consequently, the genus of $\mathcal{S}'$ is $(N-1)(qr/2-1)$.
The mapping $\gamma :(w,z)\mapsto (w,\zeta _{q}z)$ is an automorphism
of order $q$, whose fixed points can only be the $2N$ points over
$z=0,\infty $ (note that $z$ is unbranched at these points). At
the points over $z=0$ the values of $w$ are different, so these
are fixed points of $\gamma $ (which doesn't change the value of $w$).
Further, at these $N$ points $z$ is unbranched, therefore a good
local coordinate on $\mathcal{S}'$, so the definition of $\gamma $
already contains the defining relation of its monodromies at $z=0$,
which all equal $1/q$. For $z=\infty $ the value of $w$ for all
$N$ points over it is $\infty $, so they are not necessarily fixed
by $\gamma $. In order to determine the action of $\gamma $ on this
orbit, we have to consider its action in the neighborhoods of these
points. In terms of the local coordinate $\xi =1/z$ we might rewrite
\prettyref{eq:algcurv2} as \[
w^{N}=\xi ^{-q\sum _{i=1}^{r}k_{i}}\prod _{i=1}^{r}(1-(e_{i}\xi )^{q})^{k_{i}}\]
Therefore the function elements lying over $\xi =0$ are\[
w_{l}=\zeta _{N}^{l}\xi ^{-q\sum _{i=1}^{r}(k_{i}/N)}\sum _{i}a_{i}\xi ^{qi}\]
(where the term $\sum _{i}a_{i}\xi ^{qi}$ represents any, at $\xi =0$
locally holomorphic, $N$th root of $\prod _{i=1}^{r}(1-(e_{i}\xi )^{q})^{k_{i}}$).
Since $\sum _{i=1}^{r}(k_{i}/N)\in \mathbb{N}$ these are meromorphic
functions of $\xi ^{q}$, so they are fixed by $\gamma $. Therefore,
the points over $z=\infty $ are also fixed points of $\gamma $.
Just as in the case of $z=0$, $\xi $ is a good local coordinate
on $\mathcal{S}'$ at these points, so the monodromy of $\gamma $
here is $-1/q$. The total branching order of $\gamma $ is then $2N(q-1)$,
hence the genus of $\mathcal{S}'/\gamma $ is $g^{*}=(N-1)(r/2-1)$.
Finally, the trace of $\gamma $ may be expressed as

\begin{eqnarray*}
Tr(\gamma ) & = & \td {(N-1)(\frac{r}{2}-1)}{\left[\frac{1}{q}\right]^{N},\left[\frac{-1}{q}\right]^{N}}
\end{eqnarray*}
This implies the congruence relation

\begin{equation}
\mathcal{D}_{(N-1)(\frac{qr}{2}-1)}+\sum _{l=1}^{q-1}\mathcal{D}_{(N-1)(\frac{r}{2}-1)}\left(\left[\frac{l}{q}\right]^{N},\left[\frac{-l}{q}\right]^{N}\right)\equiv 0\quad \textrm{mod }q\label{eq:congrel-alg2}\end{equation}
for arbitrary primes $N\ne q$, and arbitrary $2\le r\in \mathbb{N}$.

After these general examples, where we could identify only a small
subgroup of the full automorphism group, let us now turn to curves
with larger automorphism groups, namely the modular curves. Given
a finite index normal subgroup $G\triangleleft \Gamma (1)=PSL(2,\mathbb{Z})$
one may construct a surface $\mathbb{H}/G$ by identifying the points
in orbits of $G$ on the upper half-plane $\mathbb{H}$. Compactifying
this surface amounts to adding the orbits of $G$ on $\{\infty \}\bigcup \mathbb{Q}$
(the cusps). In the following we use the notation $\mathbb{H}/G$
for this compactified surface. Elements of $\mathcal{A}=\Gamma (1)/G$,
by their natural action on the orbits, are automorphisms of the surface. 

An interesting case is when $G$ is a principal congruence subgroup 

\[
\Gamma (N)=\{A\in \Gamma (1)|A\equiv \left(\begin{array}{cc}
 1 & 0\\
 0 & 1\end{array}
\right)\; \textrm{mod}N\}\]
The closed surface $\mathcal{X}(N)=\mathbb{H}/\Gamma (N)$ is called
a modular curve. Its automorphism group $\mathcal{A}=\Gamma (1)/\Gamma (N)$
is the projective special linear group over the ring of integers mod
$N$. The first interesting surface in this series is $X(7)$, which
is the Klein quartic with automorphism group $PSL(2,7)$. The properties
of its automorphisms are well known, and the arising trace formulae
are given in \cite{genverlinde}. In general, for $N\ge 7$ the genus
of the surface $\mathcal{X}(N)$ is $g_{N}=\frac{|\mathcal{A}|}{4}(\frac{1}{6}-\frac{1}{N})+1$,
where the order of the automorphism group $\mathcal{A}$ is half of
the order of $SL(2,N)$ which is a multiplicative function of $N$
given by $|SL(2,p^{r})|=p^{3r-2}(p^{2}-1)$ for $p$ prime.

Our aim is to present a method which allows to write down the trace
formulae for automorphisms of several such surfaces. First, we need
to examine the action of $\mathcal{A}=\Gamma (1)/G$ on $\mathbb{H}/G$,
and in particular to find the orbits with non-trivial stabilizer subgroups.
Suppose that an automorphism $\gamma G\in \mathcal{A}$ fixes the
point $Gx$ (where $\gamma \in \Gamma (1)$ and $x\in \mathbb{H}\bigcup \mathbb{Q}\bigcup \{\infty \}$).
This means that $\gamma Gx=Gx\Rightarrow \gamma gx=x$ (for some $g\in G$),
i.e. $x$ is fixed by some element of $\Gamma (1)$. It is known that
there are only 3 orbits of $\Gamma (1)$ on $\mathbb{H}\bigcup \mathbb{Q}\bigcup \{\infty \}$
with non-trivial stabilizer subgroups: the orbits of the points $i$,
$\infty $ and $e^{(2\pi i)/3}$, which are fixed by $S=\left(\begin{array}{cc}
 0 & 1\\
 -1 & 0\end{array}
\right)$, $T=\left(\begin{array}{cc}
 1 & 1\\
 0 & 1\end{array}
\right)$, and $ST$, respectively. On the surface $\mathbb{H}/G$ these $\Gamma (1)$
orbits correspond to three $\mathcal{A}$ orbits with stabilizer subgroups
generated by the cosets of $S$, $T$ and $ST$ - the rest of the orbits
of $\mathcal{A}$ are regular. We may use this, for example, to compute
the genus of $\mathbb{H}/G$ using the $|\mathcal{A}|$ sheeted covering
map $\pi :\mathbb{H}/G\to (\mathbb{H}/G)/\mathcal{A}\cong \mathbb{H}/\Gamma (1)\cong \mathbb{P}^{1}$ (with e.g. $\mathbb{H}/\Gamma (1)$ denoting the compactified surface).
This also allows us to determine the signature of any element of $\mathcal{A}$
using only group theory - we shall illustrate this on an example.
Consider the surface $\mathcal{X}(11)$ of genus 26, with automorphism
group $PSL(2,11)$ whose order is 660. The coset of $T$ is of order
11 since $\left(\begin{array}{cc}
 1 & 1\\
 0 & 1\end{array}
\right)^{11}=\left(\begin{array}{cc}
 1 & 11\\
 0 & 1\end{array}
\right)\equiv \left(\begin{array}{cc}
 1 & 0\\
 0 & 1\end{array}
\right)$ mod 11. Therefore the length of the \noun{$\mathcal{A}$} orbit
of $G\infty $ (whose stabilizer subgroup is $\left\langle T\right\rangle $)
is 660/11=60. All of its points are fixed by some subgroup conjugate
to $\left\langle T\right\rangle $, and since there are 12 different
such subgroups, 60/12=5 among these points are fixed by $T$. Since
$T$ is not conjugate to $S$ or $ST$, these are all the points fixed
by $T$, whose signature is then $(g^{*}=1;11,11,11,11,11)$ (where
$g^{*}$ is computed using \prettyref{eq:hurwitz}). Similar considerations
allow us to determine the signature of any automorphism constructed
this way. 

The monodromies $k_{1}/n_{1},\ldots ,k_{r}/n_{r}$, however, depend
on the complex structure of these surfaces near the fixed points,
and may not be determined in a purely algebraic way. One possible
numerical way to get around this difficulty, if we know the signature
of the automorphism, is the use of generalized Verlinde formulae.
For all possible choices of $k_{1},\ldots ,k_{r}$ one may compute
the value of the corresponding twisted dimension in any suitable RCFT,
and check e.g. if the congruence relation \prettyref{eq:congrel}
holds. This turns out to severely restrict the possible choices of
monodromies. To illustrate this, let us again consider the automorphism
$T$ on the surface $\mathcal{X}(11)$. In this case $T$, $T^{3}$,
$T^{4}$, $T^{5}$ and $T^{9}$ all fall into one conjugacy class,
and the rest of them in another. Therefore $Tr(T)$ has to be invariant
under Galois transformations $\sigma _{l}$ for $l=3,4,5,9$, i.e.
it has to be an element of $\mathbb{Q}[\zeta _{11}+\zeta _{11}^{3}+\zeta _{11}^{4}+\zeta _{11}^{5}+\zeta _{11}^{9}]$.
It is enough to check this property for the twisted dimensions coming
from the Ising model, to find out that the only possible choices are
$\{k_{i}|i=1\ldots 5\}=\{1,1,1,2,6\}$ or $\{1,3,4,5,9\}$, plus the
sets obtained by simultaneously multiplying the elements of these
by any number mod 11. One might hope that checking other properties
(like the congruence relation), or considering other RCFT-s would
rule out some of these, but this is not the situation. There are two
reasons for this. First, recall that we have found these sets by requiring
invariance of twisted dimensions under Galois transformations $\sigma _{l}$
for $l=3,4,5,9$, so the simultaneous multiplication of $k_{i}$-s
by 3, 4, 5 or 9 doesn't change the value of the twisted dimension.
The multiplication by 2, 6, 7, 8 or 10 (or the corresponding Galois
transformation) could change the value of the twisted dimension -
this amounts to exchanging the characters corresponding to the two
conjugacy classes containing the automorphisms $T^{l}$. However,
there is an outer automorphism of $PSL(2,11)$ which exchanges the
same conjugacy classes, therefore all our constraints (which are based
on the fact that these twisted dimensions form a character of $PSL(2,11)$)
are also satisfied by the character with the twisted dimensions exchanged.
This explains the ambiguities up to simultaneous multiplication, and
in a sense this reflects our freedom in choosing which automorphism
we call $T$. The source of the remaining ambiguity is unclear, and
it leads to an interesting phenomenon. After checking in several RCFT-s,
one may conjecture that the equality\begin{equation}
\td {g}{\frac{1}{11},\frac{1}{11},\frac{1}{11},\frac{2}{11},\frac{6}{11}}=\td {g}{\frac{1}{11},\frac{3}{11},\frac{4}{11},\frac{5}{11},\frac{9}{11}}\label{eq:strangeeq}\end{equation}
 holds independently of the genus $g$. In general, similar equalities
can be found for other surfaces, preventing us from an unambiguous
determination of the monodromies.%
\footnote{ In the present case we might exclude the set $\{1,1,1,2,6\}$ by
the use of Eichler Trace Formula \cite{farkas-kra}. It is known that
the automorphism group is represented on the space of holomorphic
differentials. The Eichler Trace Formula states that the trace of
$\gamma $ in this representation can be expressed in terms of its
monodromies as \[
Tr(\gamma )=1+\sum _{i=1}^{r}\frac{exp(2\pi i\frac{k_{i}}{n_{i}})}{1-exp(2\pi i\frac{k_{i}}{n_{i}})}\]
For the choice $\{k_{i}|i=1\ldots 5\}=\{1,1,1,2,6\}$ the value of
the above expression doesn't fall in $\mathbb{Q}[\zeta _{11}+\zeta _{11}^{3}+\zeta _{11}^{4}+\zeta _{11}^{5}+\zeta _{11}^{9}]$,
but as we previously saw, this would be required from the character
of the conjugacy class of $T$ in $PSL(2,11)$. %
} Most of these equalities (like the above) don't follow from known
symmetries of twisted dimensions. However, the existence of such equalities
is not surprising. Recall that the method of writing down the trace
formula for a mapping class $\gamma $ depended on the choice of its
fixed points. Therefore, in general there can be several twisted dimensions
expressing the trace of the same mapping class, which could prove
an equality like \prettyref{eq:strangeeq} for a certain genus. For
the rest of the conjugacy classes of $PSL(2,11)$ one may unambiguously
determine the monodromies, and the results are summarized in Table
\ref{cap:psl11}.%
\begin{table}[H]
\begin{center}\begin{tabular}{|c|c|c|c|}
\hline 
order&
number&
representative&
trace\\
\hline
\hline 
&
&
&
\\
5&
264&
$\left(\begin{array}{cc}
 2 & 0\\
 0 & 6\end{array}
\right)$&
$\tdn {6}$\\
&
&
&
\\
\hline 
&
&
&
\\
11&
60&
$T,T^{3},T^{4},T^{5},T^{9}$&
$\td {1}{\frac{1}{11},\frac{3}{11},\frac{4}{11},\frac{5}{11},\frac{9}{11}}$\\
&
&
&
\\
\hline 
&
&
&
\\
11&
60&
$T^{2},T^{6},T^{7},T^{8},T^{10}$&
$\td {1}{\frac{2}{11},\frac{6}{11},\frac{7}{11},\frac{8}{11},\frac{10}{11}}$\\
&
&
&
\\
\hline 
&
&
&
\\
2&
55&
$S$&
$\td {12}{\left[\frac{1}{2}\right]^{6}}$\\
&
&
&
\\
\hline 
&
&
&
\\
3&
110&
$ST,(ST)^{2}$&
$\td {8}{\frac{1}{3},\frac{1}{3},\frac{2}{3},\frac{2}{3}}$\\
&
&
&
\\
\hline 
&
&
&
\\
6&
110&
$x^{2}=ST,y^{3}=S$&
$\td {4}{\frac{1}{2},\frac{1}{2},\frac{1}{3},\frac{2}{3}}$\\
&
&
&
\\
\hline
\end{tabular}\end{center}

\caption{\label{cap:psl11}Trace formulae for $Aut(\mathcal{X}(11))=PSL(2,11)$}
\end{table}

The resulting congruence relations are the following

\begin{eqnarray}
\tdn {26}-\tdn {6} & \equiv  & 0\quad \textrm{mod }5\nonumber \\
\mathcal{D}_{26}+5\mathcal{D}_{1}\left(\frac{1}{11},\frac{3}{11},\frac{4}{11},\frac{5}{11},\frac{9}{11}\right)+5\mathcal{D}_{1}\left(\frac{2}{11},\frac{6}{11},\frac{7}{11},\frac{8}{11},\frac{10}{11}\right) & \equiv  & 0\quad \textrm{mod }11\nonumber \\
\tdn {26}-\td {12}{\left[\frac{1}{2}\right]^{6}} & \equiv  & 0\quad \textrm{mod }2\label{eq:congrel-x11}\\
\tdn {26}-\td {8}{\frac{1}{3},\frac{1}{3},\frac{2}{3},\frac{2}{3}} & \equiv  & 0\quad \textrm{mod }3\nonumber \\
\tdn {26}+2\td {4}{\frac{1}{2},\frac{1}{2},\frac{1}{3},\frac{2}{3}}+2\td {8}{\frac{1}{3},\frac{1}{3},\frac{2}{3},\frac{2}{3}}+\td {12}{\left[\frac{1}{2}\right]^{6}} & \equiv  & 0\quad \textrm{mod }6\nonumber 
\end{eqnarray}

As another example take $G=\{A\in \Gamma (1)|A\equiv \left(\begin{array}{cc}
 2n+1 & 0\\
 0 & 2n+1\end{array}
\right)\; \textrm{mod }8,n\in \mathbb{Z}\}$=$\Gamma (8)/\left\langle \left(\begin{array}{cc}
 5 & 0\\
 0 & 5\end{array}
\right)\right\rangle $. The arising surface is $\mathcal{X}(8)/\left(\begin{array}{cc}
 5 & 0\\
 0 & 5\end{array}
\right)$ of genus $3$. Its automorphism group $(\Gamma (1)/\Gamma (8))/\left\langle \left(\begin{array}{cc}
 5 & 0\\
 0 & 5\end{array}
\right)\right\rangle $ is of order 96. The monodromies may be determined unambiguously,
and the values of all involved twisted dimensions must be integers.
The results are listed in Table \ref{cap:psl8/z2}.%
\begin{table}[H]
\begin{center}\begin{tabular}{|c|c|c|c|}
\hline 
order&
number&
representative&
trace\\
\hline
\hline 
&
&
&
\\
8&
24&
$T^{2k+1}$&
$\td {0}{\frac{1}{8},\frac{5}{8},\frac{1}{4}}$\\
&
&
&
\\
\hline 
&
&
&
\\
4&
6&
$T^{2},T^{6}$&
$\td {0}{\left[\frac{1}{4}\right]^{4}}$\\
&
&
&
\\
\hline 
&
&
&
\\
2&
3&
$T^{4}$&
$\td {1}{\left[\frac{1}{2}\right]^{4}}$\\
&
&
&
\\
\hline 
&
&
&
\\
2&
12&
$S$&
$\td {1}{\left[\frac{1}{2}\right]^{4}}$\\
&
&
&
\\
\hline 
&
&
&
\\
3&
32&
$ST,ST^{2}$&
$\td {1}{\frac{1}{3},\frac{2}{3}}$\\
&
&
&
\\
\hline 
&
&
&
\\
4&
18&
$x^{2}=T^{4}$&
$\td {1}{\frac{1}{2},\frac{1}{2}}$\\
&
&
&
\\
\hline
\end{tabular}\end{center}

\caption{\label{cap:psl8/z2}Trace formulae for $Aut\left(\mathcal{X}(8)/\left(\protect\begin{array}{cc}
 5 & 0\protect\\
 0 & 5\protect\end{array}
\right)\right)$}
\end{table}
 The arising congruence relations are\begin{eqnarray}
\tdn {3}+4\td {0}{\frac{1}{8},\frac{5}{8},\frac{1}{4}}+2\td {0}{\left[\frac{1}{4}\right]^{4}}+\td {1}{\left[\frac{1}{2}\right]^{4}} & \equiv  & 0\quad \textrm{mod }8\nonumber \\
\tdn {3}+2\td {0}{\left[\frac{1}{4}\right]^{4}}+\td {1}{\left[\frac{1}{2}\right]^{4}} & \equiv  & 0\quad \textrm{mod }4\nonumber \\
\tdn {3}+\td {1}{\left[\frac{1}{2}\right]^{4}} & \equiv  & 0\quad \textrm{mod }2\label{eq:congrel-x8}\\
\tdn {3}-\td {1}{\frac{1}{3},\frac{2}{3}} & \equiv  & 0\quad \textrm{mod }3\nonumber \\
\tdn {3}+\td {1}{\frac{1}{2},\frac{1}{2}} & \equiv  & 0\quad \textrm{mod }2\nonumber 
\end{eqnarray}
We have chosen this surface because of the remarkable property that
not only its genus, but also the trace identities for $S$ and $ST$
are the same as in the case of the Klein quartic. A possible reason
for this could be if the mapping classes of these automorphisms are
conjugate in $M_{3,0}$.

We have seen that generalized Verlinde formulae provide a connection
between Riemann surface automorphisms and arithmetic properties of
twisted dimensions. The existence of a Riemann surface automorphism
with given signature and monodromies implies (among others) a congruence
relation in the form of \prettyref{eq:congrel} and the Galois transformation
property \prettyref{eq:galoistrans}. In the case of algebraic curves
with suitable automorphisms this leads to the series of congruence
relations \prettyref{eq:congrel-alg} and \prettyref{eq:congrel-alg2}.
The first of these contains a congruence relation for any possible
signature with $g=0$, and monodromies allowed by the monodromy theorem,
while the second contains examples for signatures with $g>0$. These
relations give restrictions on the properties of allowed modular representations,
and they might be useful in classification attempts. The examples
of modular curves illustrate how generalized Verlinde formulae can
be used to obtain information about automorphic properties of Riemann
surfaces. The arising trace formulae in Tables \ref{cap:psl11}, \ref{cap:psl8/z2}
are examples where the full character of a large automorphism group
can be expressed in terms of twisted dimensions.

\section*{Acknowledgements}

I would like to thank P. Bantay for helpful discussions.

\end{document}